# The failure of the master equation for the reactive systems

Maria K. Koleva
Institute of Catalysis, Bulgarian Academy of Sciences,
1113 Sofia, Bulgaria
e-mail: mkoleva@bas.bg

**Abstract**

Two crucial for the breakdown of the master equation arguments are put forward. The first one is related to the violence of a fundamental requirement to the notion of state (thermodynamical) variable, namely: a state variable is defined provided it is insensitive to the particularities of the spatio-temporal configurations upon which the averaging over the dynamical variables proceeds. The second one is related to a ubiquitous divergence of the scattering length in the low-energy limit. In turn, it makes the rates of all the elementary processes divergent as well. Though radically novel viewpoints to the low-energy limit and to the evolution ensure the boundedness of the rates and hold the notion of a state variable available, the master equation remains inappropriate.

**Introduction**

A fundamental conjecture of the statistical mechanics is that upon averaging over a set of initial conditions, the system evolves in an ergodic manner and requires a probabilistic description. The concept implies also the existence of two types of variables: the dynamical ones, associated with the dynamical motion of the individual species and the state ones, associated with the evolution of the system as a whole. The above idea is manifested in the construction of the master equation:

$$\frac{dP(\vec{y},t)}{dt} = \int W(\vec{y}|\vec{y}')P(\vec{y}',t) - W(\vec{y}'|\vec{y})P(y,t)d\vec{y}' \qquad (1)$$

Eq.(1) is a linear equation for the evolution of the probability $P(\vec{y},t)$. The latter is probability that at the moment $t$ the system is in the state $\vec{y}$. $W(\vec{y}|\vec{y}')$ and $W(\vec{y}'|\vec{y})$ are the averaged over the the dynamical variables rates of transition between any two states $\vec{y}$ and $\vec{y}'$. The averaging is completed by means of an appropriate for the system quantum-mechanical approach. The linearity of the eq.(1) with respect to $P(\vec{y},t)$ assumes that there are two well separated time scales: a short time scale of the evolution of the fast dynamical variables and the slow time scale associated with the evolution of the state variables. The linearity of eq.(1) indicates also that the two time scales can be treated separately - at the expense of assuming the Markovian property.

Then, when the detailed balance holds, there exists a stationary distribution $P_{eq}(\vec{y})$ and any macroscopic variable and its moments can be computed through averaging over $P_{eq}(\vec{y})$.

Let us point out that the construction of the transition rates involves averaging over the dynamical variables at *fixed* values of the state ones. This presumes that the state variables are insensitive to the details of the spatio-temporal configurations involved in the averaging. An obvious requirement for that insensitivity is the spatial boundedness of the fluctuations. Indeed, the largest size of fluctuations constitutes a finite volume $V_f$ so that for any $V_i > V_f$, the state variable $c_i$ defined at volume $V_i$ uniformly converges to a value $c$ independent of $V_i$.:

$$\lim_{V_i \to \infty} c_i = c \qquad (2)$$

A lack of limits on the fluctuation size breaks the uniformity of the convergence and introduces dependence on the details of the spatio-temporal configuration(s). In turn, the latter interferes with the premise of averaging over dynamical variables at *fixed* values of the state ones.

The above considerations give rise to the question whether there is a mechanism that ensures the boundedness of the fluctuation size. The recent papers on anomalous fluctuations in extended homogeneous systems [1-2] make the problem more complicated because the non-linearity of the transition rates yields coupling of the moments of any state variable up to infinite order. In addition, the problem is rather fundamental since the non-linear transition rates are typical for almost every reactive system.

It seems natural to suppose that the non-vanishing permanent mobility of the reacting species provides the required spatial boundedness of the fluctuations. Yet, it runs into a controversy: on the one hand, under the *apriori* supposition about smallness of the fluctuations, the reaction-diffusion coupling generates coherent structures, traveling waves, etc., i.e. structures of macroscopic correlation size. On the other hand, these structures emerge around the bifurcation points where the fluctuations become anomalous large. Moreover, certain considerations about the reaction-diffusion coupling in homogeneous systems [3] yield strong dependence of the size of the giant fluctuations on the dimensionality of the system.

The problem is severely aggravated by the recently introduced process called diffusion-induced noise [4-5]. It is driven by the enduring non-correlated mobility of the reacting species and emerges at every system with hard core repulsion. If not suppressed, it makes the notion of state variable ill-defined. Next the diffusion-induced noise is presented semi-phenomenologically in the frame of the lattice-gas approach.

The modelling of the hard-core repulsion sets the relevance of the lattice-gas approach to a large variety of many-body systems. It is supposed that regardless to the details of the reaction, the species can occupy and react only at the vertices (active sites) of the lattice; no more than one species in the ground state occupies a single active site. Hereafter, the species that are already in the ground state are called adspecies.

The lattice-gas approach implies also that free species are adsorbed on the lattice where the reaction proceeds only between already adsorbed species of required types. So, the driving mechanism of the diffusion-induced noise is presented for the adsorption since it is a step prerequisite of any reaction. Given is a species trapped in a vacant site. Its further relaxation to the ground state can be interrupted by an adspecies that arrives at the same site and most probably occupies it. Thus the adspecies violates the further trapped species relaxation at that site since no more than one species can be adsorbed at a single site. The trapped species can complete the adsorption if and only if after migration it finds another vacant site. The impact of the adspecies intervention to the trapped species probability for adsorption is twofold: first, it cannot be considered as a perturbation, since it changes the adsorption potential qualitatively, namely from attractive it becomes repulsive. That is why, that type of interaction has been called diffusion-induced non-perturbative interaction. Second, the lack of coherence between the trapping moment and the moment of adspecies arrival makes the probability for adsorption multi-valued function: each selection corresponds to a certain level of relaxation at which a diffusion-induced non-perturbative interaction happens. Therefore, the adspecies mobility brings about fundamental duality of the probability for adsorption (and of the adsorption rate correspondingly): though each selection can be computed by an appropriate quantum-mechanical approach, the establishing of a given selection is a stochastic process since it is a random choice of a single selection among all available.

Since the diffusion-induced non-perturbative interactions are local events, the non-correlated mobility of the adspecies produces a lack of correlation among the established selections at any distance and at any instant. As a result, the produced adlayer would be always spatially non-homogeneous even in the academic case of identical adsorption and mobility properties of all sorts of adspecies. Moreover, the induced non-homogeneity is permanently sustained by the lack of coherence between the trapping moments and the adspecies mobility.

An immediate outcome is that the induced non-homogeneity is spread over every scale of the system. This, however, makes the notion of state variable ill-defined.

Furthermore, since the local fluctuations are unbounded and develop independently one from another, they would produce permanently non-correlated local defects and eventually would cause the system breakdown. So, the long-term stability calls for a mechanism that suppresses the induced non-homogeneity. Such mechanism has been introduced recently [6]. It is grounded on a completely new viewpoint to the coupling adlayer-lattice. The prerequisite of the novelty is the failure of the weak-coupling approach in the elimination of the induced non-homogeneity. Another prerequisite for a new approach is a ubiquitous divergence of the scattering length in the low-energy limit. It is a result of the fine-tuning caused by the everlasting fluctuations of the adsorption (reaction) potential produced by the non-correlated finite mobility of the adspecies. Correspondingly the rates of all the elementary processes become divergent as well. The new approach gives rise to a successful elimination of the induced non-homogeneity and ensures the boundedness of the amplitude of the rates. Arguments about the above considerations along with a brief outline of the approach developed in [6] are presented in the next section.

Our next task is to find out whether the proposed approach to the coupling adlayer-lattice is in agreement with the master equation. In the next section several arguments for their discrepancy are pointed out. One of them is that the elimination of the induced non-homogeneity is achieved through a global coupling of the local fluctuations *without* involving any averaging protocol.

Further, any long-range coupling of the fluctuations interferes with the fundamental premise of the statistical mechanics that every system that exerts short-ranged interactions can be divided into cells whose fluctuations are independent one from another. It should be stressed that this concept makes plausible the lack of correlation in the motion of the adspecies. Yet, the spatial coherence is necessary for sustaining long-term stability of any extended system. It is an aspect of the basic relation between the boundedeness of fluctuations and the long-term stability of the system that states: a system stays stable if the fluctuations that it exerts do not exceed its local and global thresholds of stability. Indeed, if the local fluctuations are unbounded in amplitude and in size and develop independently one from another, they would produce permanently non-correlated local defects and eventually would cause the system breakdown.

The existence of spatial coherence and the failure of the master equation requires a new viewpoint to the evolution. That is why, we have introduced [7] an approach to the motion in the state space whose novelty is that it is founded on the notion of the boundedness in its relation to the long-term stability. Its major assumption is that the boundedness of the rates renders the boundedness of the increments of the state variables. In turn, the boundedness of the increments meets the general requirement that the distance between the successive states through which a system evolves is limited so that the system stays permanently within its thresholds of stability. The incremental boundedness gives rise to strong chaotic properties of the state space. In turn, as illustrated in sec. 2, they guarantee the spatial boundedness of the fluctuations and thus justifies the notion of the state variable.

Further, the incremental boundedness gives rise to one more argument about the failure of the master equation.

### 1. Divergence of the rates. Spatial coherence of the local fluctuations

The induced non-homogeneity emerges from the lack of correlation among the fluctuations of the local rates that come from undergoing of diffusion-induced non-perturbative interactions. So, a successful mechanism that suppresses it should couple the fluctuations so that eventually all the species even their rates. Coupling of the fluctuations is an essentially non-local event and thus requires the involvement of spatially extended excitations. Since the interactions among reacting species are short-ranged, the only available non-local excitations are the cooperative excitations of the lattice (surface, interface). A successful coupling needs a feedback that acts toward evening of the initially non-identical rates making the coupled species "response" to further perturbations coherent. It should be grounded on a strong coupling adlayer-lattice, namely: the energy of colliding species dissipates to local cooperative excitations of the lattice. In turn, the impact of these local modes on the colliding species is supposed large enough to induce a new transition that dissipates through the excitation of another local cooperative modes and so on. The feedback ceases its action whenever the colliding species response becomes coherent.

However, the used so far weak coupling approach to the interaction adspecies-lattice renders any feedback local and non-correlated both in space and time. It provides only local evening of the current states of the colliding species. Indeed, suppose that the collision energy dissipates through the excitement of local cooperative modes. The weak-coupling approach considers the impact of these excitations on the Hamiltonian a perturbation. Consequently, the latter cannot produce a new transition. As a result, the interaction colliding species-lattice stops.

Along with the above argument, the necessity of a new approach to the coupling adlayer-lattice is supported by the following consideration. Next it is elucidated how the enduring non-correlated mobility of the adspecies makes the scattering length divergent in the low-energy limit. In turn, it yields divergence of the cross-section and correspondingly the rates of all the elementary processes. To unravel this point let us start with recalling that at each elementary process a species changes it state from a bounded one to a free (reaction, desorption) one or vice versa (adsorption). So, each process involves crossing of the scattering threshold. On the other hand, near the scattering threshold, the impact of the adspecies on the reaction (adsorption) potential is large enough to cause fine-tuning of a highly excited state to the scattering threshold. Since the adspecies mobility is enduring and non-correlated, the adlayer configuration permanently varies. Consequently, it produces everlasting fluctuations of the adsorption (reaction) potential that make the fine-tuning a most probable phenomenon. This immediately renders the scattering length permanently divergent. Furthermore, the divergence is rather ubiquitous than exotic phenomenon since the low-energy limit exhibits universal properties in the following sense: at low enough energies the de Broglie wavelength becomes larger than the range of interaction. This makes the scattering insensitive to the inherent structure of the scatters [8]. Thus, the universality of the low-energy limit renders the divergence of the scattering length insensitive to the chemical identity of the species.

Outlining, the above considerations call for a new approach to the coupling adlayer-lattice particularly in its low-energy limit. It imposes the following requirements: insensitivity to the chemical identity of the species and the particularities of the lattice; ubiquity - insensitivity to the details of the reaction, surface and the values of the external constraints. Next, it should ensure permanent boundedness of the rates of all the elementary processes. And last but not least, it should give rise to a successful coupling mechanism that yields suppressing of the induced non-homogeneity.

To meet the above requirements we have introduced two major assumptions [6]. The first one is that any bounded Hamiltonian separates into a "rigid" and a "flexible" part under the disturbances induced by the non-correlated mobility of the adspecies. The "rigid" part is specific to the system - it comprises those low-excited states where the impact of the mobile adspecies can be treated as perturbation. On the contrary, the "flexible" parts are associated with those highly excited states where the impact of the mobile disturbances are large enough to be treated as perturbations. Further in [6], it has been proven that in the low-energy limit the "flexible" part of the Hamiltonian exhibits strong chaotic properties. In turn, the chaoticity supports the universality of the low-energy limit since the spectrum of a chaotic Hamiltonian is characterized by a single parameter regardless to the chemical identity of the species.

The chaotic part is supposed sensitive to the environment: the dissipation of any transition energy excites cooperative modes that in turn participate to the chaotic part. This causes a change of the chaotic part so that to induce a new transition. The next assumption is that the induced transitions are non-radiative and always dissipate through excitation of appropriate cooperative modes. Thus, our two assumptions give rise to a non-perturbative feedback between the species in the chaotic states and the lattice whose most prominent property is that it "works" permanently until the global "response" to further perturbations becomes coherent.

In turn, the coherent response sustains the boundedness of the scattering length since it is a process that prevents the fine-tuning. In addition, the transitions involved in the feedback do not contribute to the cross-sections. In turn, this keeps the cross-sections finite because the transitions involved in the feedback happens in the chaotic part of the spectrum where the energies are very small. It should be stressed, that a transition that involves a vanishingly small energy makes the cross section anomalous large. Thus, the coherent response and the lack of contribution of the far infrared transitions are the warrants of the cross-sections boundedness.

Further in [6] it is shown that the collisions between the species in chaotic states, i.e. the weakest perturbations that drive the feedback, dissipate through the excitation of acoustic phonons. The particular property of the feedback created on the dissipation through acoustic phonons (gapless modes) is that the coupling area continuously enlarges on decreasing of the energy that drives the feedback.

The universality of the feedback chaotic states $\Leftrightarrow$ acoustic phonons is set on the insensitivity of the chaotic spectrum to the chemical identity of the reacting species (a chaotic spectrum is characterized by a single parameter) and the insensitivity of the acoustic phonons to the details of the adspecies configuration and the lattice itself (its dispersion relation involves a single parameter - sound velocity).

The feedback chaotic states $\Leftrightarrow$ acoustic phonons "couples" the species so that all of them share the same rate. This rate, called global rate, always equals the individual rate that comes from certain local configuration. It has been proven that the feedback always selects that individual rate which initially is in the most favorable local configuration: such that the difference in the state of that species and its immediate neighbors is the smallest. It turns out that the established throughout the system global rate does not depend on the details of the spatio-temporal configuration of the "chaotic" species and the adspecies configuration. It is proven as well that the coupling is a scale-free process that does not blur the individual properties of the "most favourable" rate.

After the evening of the local rates is completed, the relaxation continues through the "rigid" part of the spectrum where the specific properties of the system emerge. It is worth noting that the weak-coupling approach is available at the "rigid" part of the spectrum.

The most pronounced differences among the local rates come from the undergoing of a diffusion-induced non-perturbative interaction. Thus, the particularity of the "most favourable" rate is determined by the established selection. Furthermore, the lack of correlation between the trapping moments and the adspecies mobility makes all the selections equiprobable. Therefore, at next coupling session the "most favorable" rate involves another selection of the local rate. Consequently, the random choice of a single selection among all available brings about everlasting variations of the global rate in the course of time.

Summarizing, the above approach gives rise to several general properties of the global rates that make the master equation approach inappropriate, namely:

- though a current global rate always equals certain local one, it is a characteristics of the coherent behavior set by the coupling mechanism. In addition, the coupling mechanism is *not* associated with any averaging protocol. Thus, the evolution is driven by the current value of the global rates both on micro- and macro-level.
- the global rates never satisfy the detailed balance. This is a direct consequence of the duality determinism-stochasticity introduced by undergoing of a diffusion-induced non-perturbative interaction. The latter leads to a break of the time-reverse symmetry because it causes a multi-valued discontinuity of the scattering trajectories associated with different selections. The multi-valued discontinuity implies that the undergoing of a diffusion-induced non-perturbative interaction causes splitting of the scattering trajectories so that any ingoing trajectory makes a "choice" among several outgoing. It is obvious that the splitting is invariant under the time reversal. Moreover, the probability for splitting is also invariant under the time reversal because the undergoing of a diffusion-induced non-perturbative interaction is set on undergoing of a hopping from the neighborhood. Furthermore, the probability for hopping in any direction is also invariant under the time-reversal.

### 3. Chaotic kinetics and the spatial boundedness of the fluctuations

The existence of the spatial coherence and the failure of the master equation calls for a new viewpoint to the evolution. Our task is to present its grounds and that it justifies self-consistently the notion of a state variable. It has been proven [7] that if the evolution is built on the notion of boundedness in its relation to the long-term stability, the state space exhibits strong chaotic properties. The task of the present section is to prove that the chaoticity guarantees the spatial boundedness of the fluctuations and renders the notion of state variable well-defined for every state of every reactive system.

The coupling of fluctuations and the boundedness of the global rates are two aspects of the relation between the boundedness and the long-term stability, namely:

(i) the role of the spatial coherence is twofold: first, according to the introduced in [6] new approach to the coupling adlayer-lattice it is indispensable in providing permanent boundedness of the global rates. Along with it, it is necessary for sustaining long-term stability of any extended system. Indeed, if the local fluctuations are unbounded and develop independently one from another, they would produce permanently local defects and eventually would cause the system breakdown.

(ii) the permanent boundedness of the global rates renders the boundedness of the increments at the motion in the state space. In turn, the boundedness of the increments meets the general requirement that the distance between the successive states through which the system evolves is limited so that the system stays permanently within its thresholds of stability.

Indeed, it has been proven [5], that if the state variables are *apriori* defined, the boundedness of the global rates guarantees the boundedness of their increments. It is justified by the explicit relations between the concentration of the species, an intensive state variable, and the global rates for an arbitrary reaction:

$$\frac{d\vec{X}}{dt} = \vec{\alpha}\hat{A}_i(\vec{X}) - \vec{\beta}\hat{R}_j(\vec{X}) \tag{3}$$

where $\vec{X}$ is vector of the concentrations of the reaction species and the intermediates. $\hat{A}_i(\vec{X})$ is the $i-th$ selection of the global adsorption rate; $\hat{R}_j(\vec{X})$ is the $j-th$ selection of the global reaction rate. The indices serve to stress that at each moment one selection is randomly chosen among all available; $\vec{\alpha}$ and $\vec{\beta}$ are the control parameters.

To elucidate the role of the permanent variations of the global adsorption and reaction rates let us rewrite eqs.(3) in the form:

$$\frac{d\vec{X}}{dt} = \vec{\alpha}\hat{A}_{av}(\vec{X}) - \vec{\beta}\hat{R}_{av}(\vec{X}) + \vec{\alpha}\hat{\mu}_{ai}(\vec{X}) - \vec{\beta}\hat{\mu}_{ri}(\vec{X}), \tag{4}$$

where $\hat{A}_{av}(\vec{X})$ and $\hat{R}_{av}(\vec{X})$ are the mean values of the global adsorption and reaction rates at given parameter choice $\vec{\alpha}$ and $\vec{\beta}$; $\hat{\mu}_{ai}(\vec{X}) = \hat{A}(\hat{X}) - \hat{A}_{av}(\vec{X})$ and $\hat{\mu}_{ri}(\vec{X}) = \hat{R}(\vec{X}) - \hat{R}_{av}(\vec{X})$. The major property of the irregular sequences formed by $\hat{\mu}_{ai}(\vec{X})$ and $\hat{\mu}_{ri}(\vec{X})$ correspondingly is their boundedness. Consequently, the evolution of the concentration involves only bounded increments.

It has been proven [7] that a state space that obeys incremental boundedness exhibits strong chaotic properties. Since the incremental boundedness is maintained by the kinetics described by eqs.(3), the latter is called "chaotic" kinetics. It should be stressed that though resemble mean-field equations, eqs.(3)-(4) involve current values of the global rates. On the contrary, the mean-field approach involves averaged local rates.

However, the above considerations have been developed under the *apriori* assumption that the state variable(s) are defined. So, now we come to the question whether the chaotic kinetics ensures the spatial boundedness of the fluctuations. This topic is considered next.

Our present task is to illustrate that the "chaotic" kinetics provides a finite size of the fluctuations and thus guarantees the notion of a state variable well-defined.

According to the general concepts of the statistical mechanics, a fluctuation falls apart whenever its chemical potential turns to zero. Hereafter it is supposed that the fluctuations are clusters whose properties are different from the embedding "sea" but their evolution is governed by eqs.(3)-(4). The state variables are the numbers of the different sort of species involved in a cluster. It is supposed that the clusters change their size only via associating and dissociating species. So, the question becomes how the change of the number of species is interrelated to the change of the chemical potential. To reveal the problem we start with the general definition of the chemical potential $\mu$:

$$\mu_i = -\Omega \frac{\delta L}{\delta n_i} \tag{5}$$

where $\Omega$ is the volume of the system; $n_i$ is the number of the species of the $i-th$ sort; $L$ is the Langrangian.

Eq.(5) describes the thermodynamical process initiated by the small deviations from the general equilibrium condition $\delta S = 0$, where $S$ is the action.

Since the action and the Lagrangian are explicitly related to the motion in the state space, it is to be expected that the chemical potential is also explicitly related to the properties of the state space.

The further considerations are grounded on the explicit relation between the notion of chemical potential and stability of the system. Indeed, the chemical potential is a measure how strong the stability of the system is "affected" by the association or dissociation of a species. It has been found out [7] that the chaoticity of the state space is interconnected to the stability of the system in an intriguing manner. When certain relation among 3 general characteristics of the chaotic motion holds, the system is asymptotically stable (eq.(18) in [7]). In other words, the state space can be separated in the "bulk" and "surface" parts so that the "bulk" part is constituted by a "volume" whose property is that the motion inside it does not depend on the way how the trajectories approach its "boundaries". On the contrary, at the "surface" part, i.e. at the outer part of that "volume", the stability is sensitive to the way how the boundary is approached.

The above separation of the state space makes the separation of the chemical potential to a "bulk" part $\mu_b$ and a "surface" part $\mu_s$ available. Each part of the chemical potential is associated with the corresponding part of the state space. Evidently, the "bulk" chemical potential $\mu_b$ is insensitive to the change of the number of the species while the "surface" one $\mu_s$ strongly depends on it. Since any change of the number of species modifies the shape of state space "surface", the natural measure of $\mu_s$ is the local curvature of the state space "surface":

$$\mu_s = \int_S \alpha k ds \qquad (6)$$

where $\alpha$ is the density of the surface energy; $S$ is the state space surface.

The permanent variations of the number of species result in a permanent modification of the value and sign of the local curvature of the state space "surface". So, $\mu_s$ permanently varies and eventually turns the total chemical potential $\mu_{tot}$ to zero. This immediately yields the required collapse of the fluctuation.

The next task is to illustrate that this happens at finite values of the state space variables. According to the above considerations the destruction of a fluctuation happens whenever:

$$\mu_{tot} = 0 \qquad (7)$$

Since the bulk part of the chemical potential is insensitive to the variations of the state variables, eq.(7) holds when the following relation holds:

$$\mu_s = -\mu_b \qquad (8)$$

where $\mu_b$ is considered an *apriori* set constant.

Given a reaction that selects two relevant sorts of species. Hereafter their numbers are denoted by $x$ and $y$ correspondingly. The state space is two-dimensional and its "surface" can be parametrised as follows:

$$x = r^{a(\theta)} \cos\theta$$
$$y = r^{b(\theta)} \sin\theta \qquad (9)$$

where the powers $a(\theta)$ and $b(\theta)$ comprise the permanent change in the local curvature through the dependence $\theta = \theta(t)$. Then, the current local curvature $k$ reads:

$$k = \frac{\left(\dot{x}^2 + \dot{y}^2\right)^{3/2}}{\dot{x}\ddot{y} - \dot{y}\ddot{x}} \qquad (10)$$

where the derivation is with respect to the time.

Simple algebraic calculations yield:

$$k = \frac{1}{r^{2a-b}} \frac{1 - A - \dfrac{B}{r}}{\left(\sin^2\theta + r^{2(b-a)}\cos^2\theta\right)^{3/2}} \qquad (11)$$

where:

$$A = \dot{a}a\dot{b}^2 b(b-1)\sin^2\theta - \dot{b}b\dot{a}^2 a(a-1)\cos^2\theta \qquad (12)$$

$$B = \left(\dot{b}^2 + b\ddot{b}\right)\sin^2\theta + \left(\dot{a}^2 + a\ddot{a}\right)\cos^2\theta \qquad (13)$$

Since $a$, $b$ and their derivatives permanently vary, due time course eqs.(11)-(13) certainly select finite $r_{cr}$ so that eqs.(7)-(8) are satisfied. However, since the values of $a$, $b$, their derivatives and $\mu_b$ are set on the particularities of the kinetics, two cases are possible:

(I) $r_{cr} < r_{tresh}$ \qquad (14)

where $r_{tresh}$ is the size of the threshold of stability. Then, eq.(14) can be viewed as a necessary condition for a system to stay permanently stable. Besides, it can be viewed also as a necessary condition for a homogeneous system to remain permanently homogeneous.

(ii) $r_{cr} \geq r_{tresh}$ \qquad (15)

In this case the system is either destroyed or undergoes a qualitative change. The latter is viewed as a "collapse" of the old phase so that to give rise to a "birth" of a new phase. Therefore, eq.(15) can be regarded as a necessary condition not only for destabilising of the system but also for triggering processes such as nucleation, phase separation etc.

It should be stressed that the spatial boundedness of fluctuations is insensitive to the dimensionality of the system. It is set on the dimensionality of the state space. However, the latter is determined by the number of independent elementary steps involved in the reaction. Yet, the settling of a new phase strongly depends on the dimensionality of the system.

Therefore, the above considerations verify that the "chaotic" kinetics indeed ensures the spatial boundedness of fluctuations and thus renders the notion of state variable well-defined for every state of every reactive system.

Summarizing, the chaotic kinetics is self-consistent: the boundedness of the global rates give rise to the chaotic properties of the state space. In turn, they guarantee the spatial boundedness of the fluctuations and thus make the notion of the state variable available.

The incremental boundedness gives rise to one more argument about the failure of the master equation. Let us suppose that the transition from state $j$ to neighbor state $i$ depends only on whether the transition to $j$ has happened. However, the transition to $j$ depends on whether it comes from its nearest neighbourhood $k$. Hence, the transition to $i$ is set on the chain of the previous transitions …$l$…$kj$. So, is our process non-Markovian though the Chapman-Kolmogorov relation holds?! It seems Markovian because the transition from $j$ to $i$ depends only on $j$. However, it is non-Markovian, because any admissible transition depends on the succession of the previous ones. (Examples of non-Markovian chains that fulfill Chapman-Kolmogorov relation are presented in [9]). On the other hand, one of the key supposition of the master equation approach is the separation of the two time scales at the expense of assuming Markovian property.

**Conclusions**

Our considerations about the collapse of the master equation approach start with presentation of two crucial arguments. Though they affect different aspects of the master equation grounds, their origin is related to the non-correlated enduring mobility of the adspecies. Indeed, our first argument is that the notion of the state variable becomes ill-defined under the spread of the induced spatial non-homogeneity over every scale of the system. It is an outcome of the establishing of non-correlated fluctuations of the local rates whose driving "force" is the mobility of the adspecies. So, the lack of correlation in the mobility results in a lack of correlation among the selections, i.e. among the fluctuations of the local rates. Our second argument is that the enduring non-correlated mobility of the adspecies gives rise to permanent fluctuations of the adsorption (reaction) potential. In turn, the everlasting fluctuations make the fine-tuning a ubiquitous phenomenon in the low-energy limit and thus render the rates of all the elementary processes divergent.

To overcome the above difficulties, we introduce a new viewpoint to the coupling adlayer-lattice. It ensures the elimination of the induced non-homogeneity and the boundedness of the rates of all the elementary processes at the expense of coupling of the local fluctuations over the entire system. However, any long-range coupling of the fluctuations interferes with the fundamental idea of the statistical mechanics that every system that exerts short-ranged interactions can be divided into cells whose fluctuations are independent one from another. It should be stressed that this premise makes the lack of correlation in the motion of the adspecies plausible. Yet, the spatial coherence is necessary for sustaining long-term stability of any extended system. It is an aspect of the basic relation between the boundedeness of fluctuations and the long-term stability of the system that says: a system stays stable if the fluctuations that it exerts do not exceed its local and global thresholds of stability. Indeed, if the local fluctuations are unbounded and develop independently one from another, they would produce permanently local defects and eventually would cause the system breakdown.

The existence of spatial coherence and the permanent boundedness of the global rates call for a new viewpoint to the evolution. That is why, we introduce a new approach to the motion in the state space based on the notion of the boundedness. Its major assumption is that the boundedness of the rates renders the boundedness of the increments of the state variables. In turn, the boundedness of the increments meets the general requirement that the distance between the successive states is limited so that the system stays permanently within its thresholds of stability.

A major property of this approach is that it integrates self-consistently the notion of the state variable.

Outlining, we overcome the difficulties to the master equation approach at the expense of assuming new concepts. However, they also leave the master equation unsuitable.